\RequirePackage{amsmath}
\documentclass{llncs}
\usepackage{graphicx}
\usepackage{color}
\usepackage{soul}
\usepackage{enumitem}
\usepackage{hyperref}
\usepackage{url,footnote}
\usepackage{dirtytalk}
\usepackage[utf8]{inputenc}
\usepackage[english]{babel}
\usepackage{sidecap}
\usepackage{wrapfig}
\usepackage{multirow}
\usepackage[table,xcdraw]{xcolor}
\usepackage{listings}
\usepackage{amsmath}
\usepackage{amssymb}
\usepackage{stmaryrd}
\usepackage[linesnumbered,ruled,lined,boxed,commentsnumbered]{algorithm2e}
\usepackage{ifsym}
\usepackage{booktabs}
\graphicspath{{figures/}}

\definecolor{codegreen}{rgb}{0,0.6,0}
\definecolor{codegray}{rgb}{0.5,0.5,0.5}
\definecolor{codepurple}{rgb}{0.58,0,0.82}
\definecolor{backcolour}{rgb}{0.95,0.95,0.92}
 
\lstdefinestyle{mystyle}{
    backgroundcolor=\color{backcolour},   
    commentstyle=\color{codegreen},
    keywordstyle=\color{magenta},
    numberstyle=\tiny\color{codegray},
    stringstyle=\color{codepurple},
    basicstyle=\footnotesize,
    breakatwhitespace=false,         
    breaklines=true,                 
    captionpos=b,                    
    keepspaces=true,                 
    numbers=left,                    
    numbersep=5pt,                  
    showspaces=false,                
    showstringspaces=false,
    showtabs=false,                  
    tabsize=2
}
 
\hypersetup{
    colorlinks=true,
    linkcolor=blue,
    filecolor=blue,      
    urlcolor=cyan,
}
\usepackage[textsize=tiny]{todonotes} 

\begin{document}

\title{Towards an Integrated Graph Algebra for Graph Pattern Matching with Gremlin}

\author{Harsh Thakkar\inst{1}, S\"oren Auer\inst{1,2}, Maria-Esther Vidal\inst{2}}
\institute{Smart Data Analytics Lab (SDA), University of Bonn, Germany
\and TIB \& Leibniz University of Hannover, Germany\\
\email{\{lastname\}@cs.uni-bonn.de\inst{1}, dpunjani@di.uoa.gr\inst{2}}}

\maketitle

\begin{abstract}
Graph data management (also called NoSQL) has revealed beneficial characteristics in terms of flexibility and scalability by differently balancing between query expressivity and schema flexibility.
This peculiar advantage has resulted into an unforeseen race of developing new task-specific graph systems, query languages and data models, such as property graphs, key-value, wide column, resource description framework (RDF), etc.
Present-day graph query languages are focused towards flexible graph pattern matching (aka sub-graph matching), whereas graph computing frameworks aim towards providing fast parallel (distributed) execution of instructions.
The consequence of this rapid growth in the variety of graph-based data management systems has resulted in a lack of standardization.
Gremlin, a graph traversal language, and machine provide a common platform for supporting any graph computing system (such as an OLTP graph database or OLAP graph processors).
In this extended report, we present a formalization of graph pattern matching for Gremlin queries.
We also study, discuss and consolidate various existing graph algebra operators into an integrated graph algebra.
\end{abstract}

\section{Introduction}\label{sec:introduction}

Upon observing the evolution of information technology, we can observe a trend from data models and knowledge representation techniques being tightly bound to the capabilities of the underlying hardware towards more intuitive and natural methods resembling human-style information processing.
This evolution started with machine assembly languages, went over procedural programming, object-oriented methods and resulted in an ever more loosely coupling of data and code with relational data bases, declarative query languages and object-relational mapping (ORM).
In recent years, we can observe an even further step in this evolution -- graph based data models, which organize information in conceptual networks.
Graphs are valued distinctly, when it comes to choosing formalisms for modelling real-world scenarios such as biological, transport, communication and social networks due to their intuitive data model.
In particular, the property graph data model is capable of representing complex domain networks~\cite{rodriguezN11grptrvslpttrn}.

Graph analysis tools have turned out to be one of pioneering applications in understanding these natural and man-made networks~\cite{gomes2015beta}. 
Graph analysis is carried out using graph processing techniques which ultimately  boil down to efficient graph query processing. 
Graph Pattern Matching (GPM), also referred to as the sub-graph matching is the foundational problem of graph query processing. 
Many vendors have proposed a variety of (proprietary) graph query languages to demonstrate the solvability of graph pattern matching problem.

These modern graph query languages focus either on \textit{traversal}, where traversers move over vertices and edges of a graph in a user defined fashion or on  \textit{pattern-matching}, where graph patterns are matched against the graph database. 
\textit{Gremlin}~\cite{rodriguez2015trvslmchnlang} is one such modern graph query language, with a distinctive advantage over others that it offers both of these perspectives. 
This implies that a user can reap benefits of both declarative and imperative matching style within the same framework. 
Furthermore, conducting GPM in Gremlin can be of crucial importance in cases such as:
\begin{itemize} 
    \item Querying very large graphs, where a user is not completely aware of certain dataset-specific statistics of the graph (e.g., the number of \texttt{created} vs \texttt{knows} edges existing in the graph (ref. Figure~\ref{fig:property_graph}));
    \item Creating optimal query plans, without the user having to dive deep into traversal optimization strategies.
    \item In application-specific settings such as a question answering~\cite{UngerNC16QALD}, users express information needs (e.g., natural language questions) which can be better represented as graph pattern matching problems than path traversals.
\end{itemize}

\noindent 
In this work, we present an extended version of our earlier work \cite{thakkar2017graph} that contributes to establishing a formal base for a graph query algebra, by surveying and integrating existing graph query operators.
The contributions of this work are in particular:
\begin{itemize} 
    \item We consolidate existing graph algebra operators from the literature and propose two new traversal operators into an integrated graph algebra.
    \item We formalize the graph pattern matching construct of the Gremlin query language.
    \item We provide a formal specification of pattern matching traversals for the Gremlin language, which can serve as a foundation for implementing a Gremlin based query compilation engine.
\end{itemize}

As a result, the formalization of graph query algebra supports the integration and interoperability of different graph data models~\cite{DBLP:conf/amw/AnglesTT19} (e.g., executing SPARQL queries on top of Gremlin~\cite{thakkar2018stitch,91512}), helps to prevent vendor lock in scenarios and boosts data management benchmarking efforts such as LITMUS~\cite{keswani2017litmus,thakkar2017towards,DBLP:conf/i-semantics/ThakkarKDLA17} and the Linked Data Benchmark Council (LDBC)~\cite{angles2014linked}.

The remainder of this article is organised as follows: 
\autoref{sec:background} describes the preliminaries including property graphs, graph pattern matching, and foundations of relational algebra.
\autoref{sec:gremlin_gpm} elaborates on Gremlin as a traversal language and machine, and discusses synergy of GPM and its evaluation in Gremlin.
\autoref{sec:algorithm} proposes the Gremlin to graph algebra mapping algorithm. 
\autoref{sec:relwork} surveys related work. 
In \autoref{sec:conclusion} we conclude this article and discuss future work.

\section{Preliminaries}\label{sec:background}

In this section, we present and summarise the mathematical concepts used in this article. 
Our notation closely follows~\cite{holsch2016algebra} and extends~\cite{rodriguezN11pathalgebra} by adding the notion of vertex labels.     

\subsection{Property Graphs}\label{sec:pg_def}

Property graphs, also referred to as directed, edge-labeled, attributed multi-graphs, have been formally defined in a wide variety of texts, such as~\cite{angles2016foundations,gubichev2014graph,krause2016sql,rodriguezN11grptrvslpttrn,prud2006sparql}. 
We adapt the definition of property graphs presented by~\cite{rodriguezN11grptrvslpttrn}: 

\begin{definition}[Property Graph]\label{def:pg}
A property graph is defined as $G = \{V, E, \lambda, \mu\}$; where: 
\begin{itemize}
    \item $V$ is the set of vertices, 
    \item $E$ is the set of \textit{directed} edges such that $E \subseteq (V \times Lab \times V)$ where $Lab$ is the set of Labels, 
    \item $\lambda$ is a function that assigns labels to the edges and vertices (i.e. $\lambda : V \cup E \rightarrow \Sigma^{*}$)\footnote{set of strings ($\Sigma^{*}$)}, and 
    \item $\mu$ is a partial function that maps elements and keys to values (i.e. $\mu : (V \cup E) \times R \rightarrow S)$ i.e. properties (key $r \in R$, value $s \in S$). \qed
\end{itemize} 
\end{definition}
        
For simplicity, we define disparate sets (of $\mu$ and $\lambda$) for the labels and properties of vertices and edges respectively, adapting the terminology used in~\cite{holsch2016algebra}. 
We define:
\begin{itemize}
    \item $L_{v}$: Set of vertex labels $(L_{v} \subset \Sigma^{*})$, $\lambda_{l} : V \rightarrow {L_{v}}$ 
    assigns label to each vertex
    \item $L_{e}$: Set of edge labels ($L_{e} \subset \Sigma^{*})$, $\lambda_{e} : E \rightarrow {L_{e}}$ assigns label to each edge. ($L_{v} \cap L_{e} = \phi$ , $L_{v} \cup L_{e} = Lab$; wrt definition~\ref{def:pg}, $\lambda = \lambda_{v} \cup \lambda_{e}$)
    \item $P_{v}$:Set of vertex properties $(P_{v} \subset \textit{R})$, $\mu_{v} : V \times P_{v} \rightarrow S $ assigns a value to each vertex property
    \item $P_{e}$: Set of edge properties $(P_{e} \subset \textit{R})$, $\mu_{e} : E \times P_{e} \rightarrow S$ assigns a value to each edge property ($P_{v} \neq P_{e}$; in the context of definition~\ref{def:pg}, $\mu = \mu_{v} \cup \mu_{e}$)
\end{itemize}
    
\begin{figure}[tb]
    \centering
    \includegraphics[width=\textwidth]{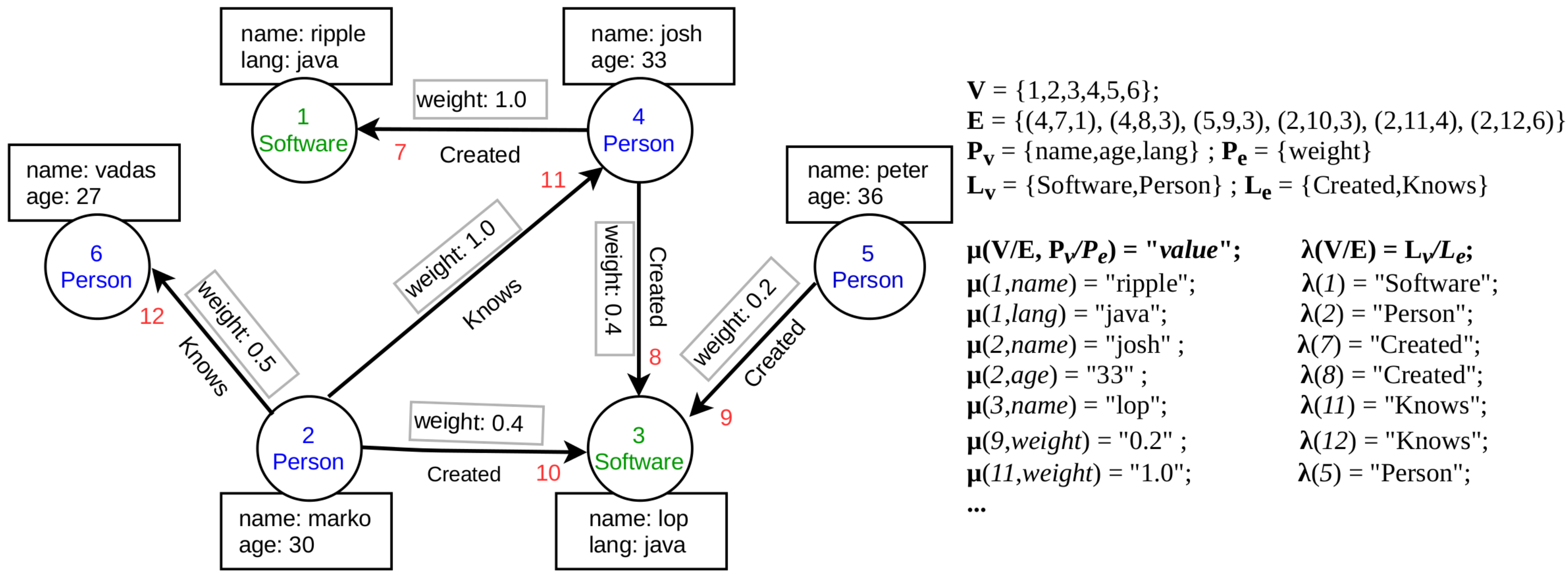}
    \caption{\textbf{Running example.} This figure presents a collaboration network scenario of employees in a typical software company, 
    There are 6 vertices (with labels "person" and "software") and 6 edges connecting them with two type of relations (i.e., edge labels) namely, "created" and "knows".}
    \label{fig:property_graph}
\end{figure}

Figure~\ref{fig:property_graph}, presents a visualization of the \textit{Apache TinkerPop} modern crew graph\footnote{TinkerPop Modern Crew property graph (\url{http://tinkerpop.apache.org/docs/3.2.3/reference/\#intro})}. 
We use this graph as a running example throughout this paper.

\subsection{Graph Pattern Matching}\label{sec:gen_gpm}
\begin{figure}[tb]
    \centering
    \includegraphics[width=\textwidth]{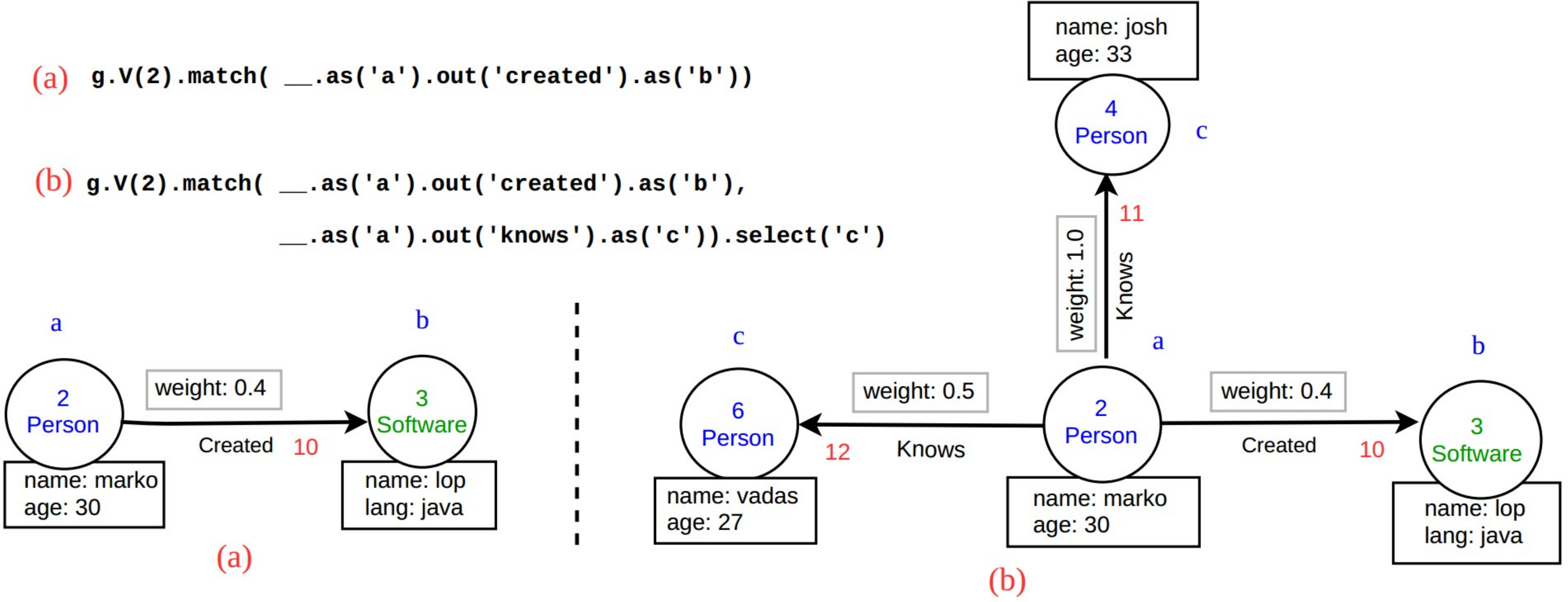} 
    \caption{We illustrate the notion of a sample, {\color{red}(a)} basic graph pattern and {\color{red}(b)} complex graph pattern, as a gremlin traversal over the graph G as shown in Figure~\ref{fig:property_graph}.}
    \label{fig:example_1}
\end{figure}

Graph Pattern Matching (GPM) is a computational task consisting of matching graph patterns (\textit{P}) against a graph (\textit{G}, ref. def.~\ref{def:pg}).
Graph databases perform GPM for querying a variety of data models such as RDF, Property Graphs, edge-labelled graphs, etc., and many works address and analyze its solvability, such as~\cite{angles2016foundations,gubichev2014graph,krause2016sql,martonformalizing,van2016pgql}.
Various graph query languages have been implemented for querying these data models, that are of imperative and declarative  nature, such as:
\begin{itemize} 
    \item The SPARQL\footnote{\url{https://www.w3.org/TR/rdf-sparql-query/}} query language for RDF triple stores (declarative), 
    \item Neo4J's native query language CYPHER\footnote{\url{https://neo4j.com/developer/cypher-query-language/}} (declarative),
    \item Apache TinkerPop's graph traversal language and machine \textit{Gremlin}\footnote{\url{https://tinkerpop.apache.org/}} (a functional language traditionally imperative, but also offers a declarative construct). 
\end{itemize}
GPM queries can represented by two types of graph patterns~\cite{angles2016foundations}, basic graph patterns (BGP) and complex graph patterns (CGP), which add operations such as projections, unions, etc. to BGPs (cf. Figure~\ref{fig:example_1}).

\begin{example}
    For the graph pattern (say \textit{P}) as shown in Figure~\ref{fig:example_1}{\color{blue}(b)}, as per the Definition~\ref{def:pg} of property graphs, we have that P = \{V, E, $\lambda$, $\mu$\}, where Figure~\ref{fig:mapping} demonstrates the values of its components.  
    \begin{figure}[ht]
        \centering
      \includegraphics[width=\textwidth]{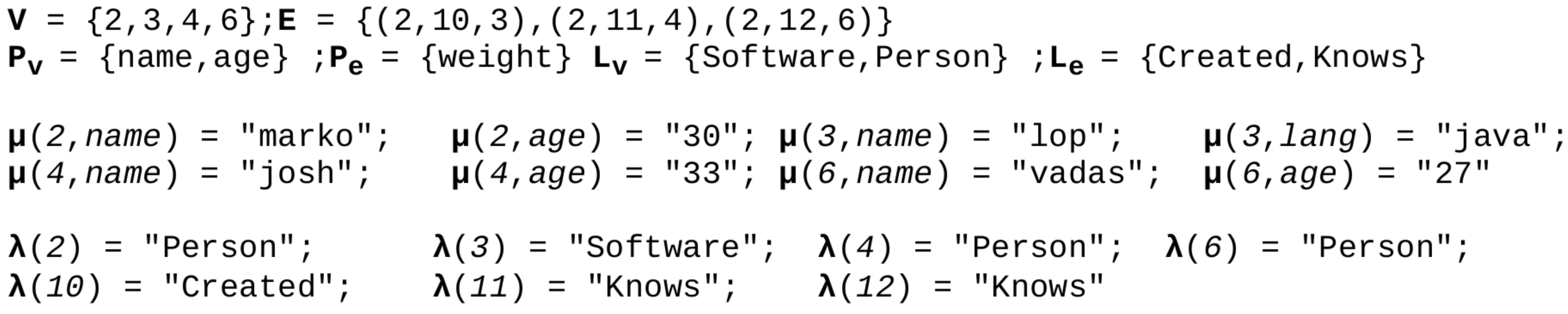} 
        \caption{Demonstrating the components of the CGP shown in Figure~\ref{fig:example_1}{\color{blue}(b)}.}
        \label{fig:mapping}
    \end{figure}
\end{example}

A GPM is evaluated by \textit{matching}, a sub-graph pattern over a graph G. 
Matching has been formally defined in various texts and we summarise a formal definition in our context which closely follows the definition provided by~\cite{gubichev2014graph,angles2016foundations}.
    
\begin{definition}[Match of a graph pattern]\label{def:match}
A graph pattern $P = (V_{p}, E_{p}, \lambda_{p}, \mu_{p})$; is matching the graph G = $(V,E,\lambda,\mu)$, iff the following conditions are satisfied:
\begin{enumerate} 
     \item there exist mappings $\mu_{p}$ and $\lambda_{p}$ such that, all variables are mapped to constants, and all constants are mapped to themselves (i.e. $ \lambda_{p} \in \lambda, \mu_{p} \in \mu$),
     \item each edge \'e $\in E_{p}$ in $P$ is mapped to an edge $e \in E$ in $G$, each vertex \'v $\in V_{p}$ in $P$ is mapped to a vertex $v \in V$ in $G$, and
     \item the structure of $P$ is preserved in $G$ (i.e. P is a sub-graph of $G$) \qed
\end{enumerate}
\end{definition}
We discuss \textit{matching} graph patterns, over a graph $G$, in the context of Gremlin traversal language in Section~\ref{sec:gremlin_match}.
    
\begin{example}
    We illustrate the evaluation of graph patterns (a) and (b) from Figure~\ref{fig:example_1}, over the graph $G$ from Figure~\ref{fig:property_graph}.
    \begin{center}
    {\color{red}(a)} \texttt{v[3]} \hspace{20pt} {\color{red}(b)} \texttt{v[4]}\footnote{Please note that since we did not explicitly specify any values, the traversal returns the ids of the "matched" elements as a result.} \\
    \hspace{75pt} \texttt{v[6]}
    \end{center} 
\end{example}

In Gremlin, GPM is performed by traversing\footnote{The act of visiting of vertices ($v \in V$) and edges ($e \in E$) in a graph in an alternating manner (in some algorithmic fashion)~\cite{rodriguezN11grptrvslpttrn}.} over a graph $G$. A traversal $t$ over $G$ derives paths of arbitrary length. 
Therefore, a GPM query in Gremlin can be perceived as a path traversal. 
Rodriguez et al.~\cite{rodriguezN11pathalgebra} define a \textit{path} as:    

\begin{definition}[Path]\label{def:path}
    A path $p$ is a sequence or a string, where \textit{p} $\in E^{*}$ and $E \subset (V \times L_{e} \times V$)\footnote{The kleene start operation * constructs the free monoid E$^{*}$ = $\bigcup_{n=0}^{\infty}$ E$^{i}$. where E$^{0}$ = \{$\epsilon$\}; $\epsilon$ is the identity/empty element.}. A path allows for repeated edges and the length of a path is denoted by $||$\textit{p}$||$, which is equal to the number of edges in \textit{p}. \qed
\end{definition}

Moreover, from~\cite{rodriguezN11grptrvslpttrn} we also know that these path queries are comprised of several atomic operations called the single-step traversals.
A list of graph-specific operators have been defined in~\cite{rodriguezN11grptrvslpttrn,rodriguezN11pathalgebra}. 
We outline these and other operators, in the following.
    
\subsection{Graph Algebra Foundations}
We present a consolidated summary of various 
graph query operators defined in the literature~\cite{AnglesG16Multiset,holsch2016algebra,martonformalizing,perez2006semantics,rodriguezN11pathalgebra,rodriguezN11grptrvslpttrn}. 
For brevity, we abstain from dwelling into rigorous formal definitions and underlying proofs and refer the interested reader to the respective articles.

\subsubsection{Unary operators} 

\paragraph{Projection} ($\pi_{a,b,..}$) : $R \cup S \rightarrow \Sigma^{*}$: operator projects values of a specific set of variables $a,b,..,n$ (i.e. keys and elements), from the solution of a matched input graph pattern $P$, against the graph $G$. 
    Moreover, the results returned by ($\pi_{a,b}$) are not deduplicated be default, i.e. the result will contain as many possible matched values or items as the input pattern $P$. 
    This operator is present in all standard graph query languages (e.g. \texttt{SELECT} in SPARQL, and \texttt{MATCH} in CYPHER).

\paragraph{Selection} ($\exists(p$)), is analogous to the filter operator ($\sigma$), as defined in~\cite{martonformalizing,angles2016foundations}, restricts the match of a certain graph pattern $P$ against a graph $G$, by imposing conditional expressions (p) e.g., inequalities and/or other traversal-specific predicates (where predicate is a proposition formula).        

\subsubsection{Binary operators}\label{sec:bin_ops}

\paragraph{Concatenation}~\cite{rodriguezN11pathalgebra} ($\circ$): $E^{*} \times E^{*} \rightarrow E^{*}$: concatenates two paths (cf. Definition~\ref{def:path}). 
For instance, if $(i,\alpha,j) \text{ and } (j,\beta,k)$ are two edges in a graph G, then their concatenation is the new path $(i,\alpha,j,\beta,k)$; where $i,j,k \in V$ and $\alpha, \beta \in L_{e}$.

\paragraph{Union} operator ($\uplus$): $P(E^{*}) \times P(E^{*}) \rightarrow P(E^{*})$ : is the multiset union\footnote{Note that the domains and ranges of each of these sets are the power sets.} (bag union) of two path traversals or graph patterns. 
For instance, \{(1,2), (3,4), (3,4), (4,5)\} $\uplus$ \{(1,2), (3,4)\} = \{(1,2), (1,2), (3,4), (3,4), (3,4), (4,5)\}. 
The results of this operator, like projection, are not deduplicated by default.

\paragraph{Join}~\cite{rodriguezN11pathalgebra} ($\bowtie_{\circ}$): $P(E^{*}) \times P(E^{*}) \rightarrow P(E^{*})$ : produces the concatenative join of two sets of paths (path traversals) such that if $P,R \in P(E^{*}$), then
    \begin{center}
        $P \bowtie_{\circ} R = \{p \circ r \mid p \in P \wedge r \in R \wedge (p = \epsilon \vee r = \epsilon \vee \gamma^{+}(p) = \gamma^{-}(r))\}$\footnote{Here, ($\gamma^{-}, \gamma^{+}$) denote the first and last elements of a path respectively.}
    \end{center}
    For instance, if $P = \{(v_{1},e_{1},v_{2}), (v_{2},e_{2},v_{3})\}$ and $R = \{(v_{2},e_{2},v_{3}), (v_{2},e_{2},v_{1})\}$, then, 
    \begin{center}
        $P \bowtie_{\circ} R = \{(v_{1},e_{1},v_{2},v_{2},e_{2},v_{3}), (v_{1},e_{1},v_{2},v_{2},e_{2},v_{1})\}$,
    \end{center} 
where $v_{1},v_{2},v_{3} \in V; e_{1},e_{2} \in L_{e}$. 
    
\paragraph{Left}-join ({\tiny \textifsym{d|><|}}), \textit{Right}-join ({\tiny{\textifsym{|><|d}}}) and the \textit{Anti}-join ($\triangleright$) operators: these operators, are not explicitly implemented in Gremlin, unlike in other graph query languages~\cite{rodriguez2015trvslmchnlang}. 
Their results can, however, be simulated by the user, at run-time via selecting desired values of elements, vertices or edges declaring using the projection operator. 
For instance, an anti-join can be "computationally" achieved by \texttt{not}'ing an argument in the match step by using \texttt{.not()} Gremlin step. 

\subsubsection{General Extensions.} We borrow the extended relational operators \textit{Grouping} ($\dagger_{a}$(p)), \textit{Sorting} ($\Re_{\Uparrow a, \Downarrow b}$(p)) and \textit{Deduplication} ($\delta_{a,b,..}$(p)) which have been defined in~\cite{martonformalizing}. 
A detailed illustration with formal definitions of extended operators can be found in~\cite{garcia2009database}. 
    
\subsubsection{Graph-specific Extensions.}\label{sec:graph-specific}
Various graph/traversal-specific operators have been defined in works such as~\cite{holsch2016algebra,rodriguezN11pathalgebra}. 
Furthermore, there also exist certain application-specific extensions of algebra operators, such as the $\alpha$ and $\beta$ operators, for graph data aggregation (used in complex graph network analysis) defined by~\cite{gomes2015beta}.  
We present graph-specific operators, some of which have been adapted from~\cite{holsch2016algebra,martonformalizing} and propose additional operators based on the algebra defined by~\cite{rodriguezN11pathalgebra,rodriguezN11grptrvslpttrn}. 
    \begin{itemize} 
        \item The \textit{Get-vertices/Get-edges} nullary operators (V$_{g}$/ E$_{g}$): return the list of vertices/edges, respectively. 
        These operators, w.r.t. Gremlin query construct, denote the start of a traversal.
        It is also possible to traverse from a specific vertex/edge in a graph, given their id's.
        Furthermore, they can be used to construct custom indexes over elements depending on user's choice. 
        \item The \textit{Traverse operator} ($\updownarrow_{v1}^{v2}${[}e{]}(p)): $P(V \cup E) \times \Sigma^{*} \rightarrow P(V \cup E)$ : is an adapted version, analogues to the \textit{expand-both} operator defined by~\cite{holsch2016algebra}. 
        The traverse operator represents the traversing over the graph operation (traversing \textit{in} $\downarrow$ or \textit{out} $\uparrow$ from a vertex (\textit{v1}) to an adjacent vertex (\textit{v2}) given the edge label \texttt{[e]}, where $(v1,v2) \in V, e \in L_{e})$.
        \item The \textit{Property filter} operator ($\sigma_{condition}^{v/e}$(p)): $ P(V \cup E) \times S \rightarrow \Sigma^{*}$ : is a binary operator which: \textit{(i)} filters the values of selected element (vertex/edge), if a \textit{condition} is declared, \textit{(ii)} otherwise, it simply returns the value of the element's property.
        \item The \textit{Restriction} unary operator ($\lambda_{l}^{s}$(p)) is an adaptation of~\cite{li2005ranksql}, which we borrow from~\cite{martonformalizing}. 
        It takes a list as input and returns the top \textit{s} values, skipping specified \textit{l} values. 
        It is analogous to the LIMIT and OFFSET modifier keyword pair in SPARQL.
    \end{itemize}

\begin{table}[tb]
    \centering
    \begin{tabular}{||c|c|c|c|c||}
    \rowcolor[HTML]{656565} \hline
    {\color[HTML]{FFFFFF} \textbf{Ops}} & {\color[HTML]{FFFFFF} \textbf{Operation}} & {\color[HTML]{FFFFFF} \textbf{Operator}} & {\color[HTML]{FFFFFF} \textbf{Gremlin Step}} & {\color[HTML]{FFFFFF} \textbf{Step type}} \\ \hline \hline
     & Get vertices & V$_{g}$ & \texttt{g.V()} & - \\
    \multirow{-2}{*}{0} & Get edges & E$_{g}$ & \texttt{g.E()} & - \\ \hline \hline
     & Selection & $\exists$(p) & \texttt{.where()} & Filter \\
     & Property filter & $\sigma_{condition}^{a}$(p) & \texttt{.has()/.values()} & Filter \\
     & Projection & $\Pi_{a,b,...}$(p) & \texttt{.select()} & Map \\
     & De-duplication & $\delta_{a,b,..}$(p) & \texttt{.dedup()} & Filter \\
     & Restriction & $\lambda_{l}^{s}$(p) & \texttt{.limit()} & Filter \\
     & Sorting & $\Re_{\Uparrow a, \Downarrow b}$(p) & \texttt{.order().by()} & Map \\
     & Grouping & $\dagger_{a}$(p) & \texttt{.group().by()} & Map/SideEffect \\
    \multirow{-8}{*}{1} & Traverse (out/in) & $\updownarrow_{v1}^{v2}${[}e{]}(p) & \texttt{.out()}/\texttt{.in()} & FlatMap \\ \hline \hline
     & Join & p $\bowtie_{\circ}$ r & \texttt{.and()} & Filter \\
    \multirow{-2}{*}{2} & Union & p $\uplus$ r & \texttt{.union()} & Branch \\ \hline
    \end{tabular} 
    \caption{A consolidated list of relational algebra graph operators with their corresponding instruction steps in Gremlin traversal language.} 
    \label{tab:ops_master}
\end{table}
   
\section{The Gremlin Graph Traversal Language and Machine}\label{sec:gremlin_gpm}

Gremlin is the query language of Apache TinkerPop\footnote{Gremlin: Apache TinkerPop's graph traversal language and machine~(\url{https://tinkerpop.apache.org/})} graph computing framework. 
Gremlin is system agnostic, and enables both -- pattern matching (declarative) and graph traversal (imperative) style of querying over graphs. 

\subsubsection{The Machine.} Theoretically, a set of traversers in \textit{T} move (traverse) over a graph G (property graph, cf. Section~\ref{sec:pg_def}) according to the instruction in ($\Psi$), and this computation is said to have completed when there are either: 
\begin{enumerate}
    \item no more existing traversers ($t$), or 
    \item no more existing instructions ($\psi$) that are referenced by the traversers (i.e. program has halted).
\end{enumerate}
Result of the computation being either a null/empty set (i.e. former case) or the multiset union of the graph locations (vertices, edges, labels, properties, etc.) of the halted traversers which they reference. Rodriguez et al.~\cite{rodriguez2015trvslmchnlang} formally define the operation of a traverser \textit{t} as follows: 
\begin{equation}
    G \leftarrow \hspace{2pt} _{\mu}  \frac{t \in T}{\{\beta, \varsigma\}}  \hspace{2pt} _{\psi} \rightarrow \Psi 
\end{equation}
where, $\mu$: \textit{T} $\rightarrow$ \textit{U} is a mapping from the traverser to its location in G; $\psi$: \textit{T} $\rightarrow$ $\Psi$ maps a traverser to a step in $\Psi$; $\beta$: \textit{T} $\rightarrow$ $\mathbb{N}$ maps a traverser to its bulk\footnote{ The bulk of a traverser is number of equivalent traversers a particular traverser represents.}; $\varsigma$: \textit{T} $\rightarrow$ \textit{U} maps a traverser to its sack (local variable of a traverser) value. 

\subsubsection{The Traversal.} A Gremlin graph traversal can be represented in any host language that supports function composition and function nesting. 
These steps are are either of: 
\begin{enumerate}
    \item \textit{Linear motif} - $ f \circ g \circ h $, where the traversal is a linear chain of steps; or
    \item \textit{Nested motif} - $ f \circ ( g \circ h) $ where, the nested traversal $ g \circ h $ is passed as an argument to step \textit{f}~\cite{rodriguez2015trvslmchnlang}. 
\end{enumerate}
A step ($ f \in \Psi $) can be, defined as $ f : A^{\star} \rightarrow B^{\star} $\footnote{The Kleene star notation ($A^{\star}, B^{\star}$) denotes that multiple traversers can be in the same element (A,B).}. 
Where, \textit{f} maps a set of traversers of type A (located at objects of A) to a set of traversers of type B (located at objects of B).
Given that Gremlin is a language and a virtual machine, it is possible to design another traversal language that compiles to the Gremlin traversal machine (analogous to how Scala compiles to the JVM).
As a result, there exists various Gremlin dialects such as Gremlin-Groovy, Gremlin-Python, etc. 

A Gremlin traversal ($\Psi$) can be compiled down to a collection of steps or instructions which form the Gremlin instruction set.
The Gremlin instruction set comprises approximately 30 steps which are sufficient to provide general purpose computing and for expressing the common motifs of any graph traversal query, as highlighted by~\cite{rodriguez2015trvslmchnlang}.
However, in a majority of cases only around 10 of these instructions are sufficient to address the most common information needs (i.e. for graph pattern matching and traversal).

\subsection{Graph Pattern Matching Queries in Gremlin}\label{sec:gremlin_gpmq}
Gremlin provides the GPM construct, analogous to SPARQL~\cite{AnglesG16Multiset,perez2006semantics,schmidt2010foundations}\footnote{However, Property Graphs do not encode all semantics of RDF Graphs (e.g. blank nodes.}, using the \texttt{Match()}-step. 
This enables the user to represent the query using multiple individual connected or disconnected graph patterns. 
Each of these graph patterns can be perceived as a simple path traversal, to-and-from a specific source and destination, over the graph. 

Each traversal is a path query starting at a particular source (A) and terminating at a destination (B) by visiting vertices (v $\in$ V) and edges (e $\in$ E (V $\times$ V) ) in an alternating fashion (i.e. referred to as traversing~\cite{rodriguezN11grptrvslpttrn}). 
Each path query is composed of one or more single-step traversals. 
Through function composition and currying , it is possible to define a query of arbitrary length~\cite{rodriguezN11grptrvslpttrn}. 
These path queries can be a combination of either a source, destination, labelled traversal or all of them in a varying fashion, depending on the query defined by the user. 

\begin{example}
For instance, consider a simple path traversal to the oldest person that marko knows over the graph G as show in Figure~\ref{def:pg}. 
Listing~\ref{lst1} represents the gremlin query for the described traversal.
\begin{lstlisting}[caption= {Return the age of the oldest person marko knows}, label={lst1}, language=Java]
g.V().has("name","marko").out("knows").values("age").max()
\end{lstlisting}
Here, \texttt{g.V()} i.e. $V_{g}$ is the traverser definition bijective to \textit{V} where, $\uplus_{i} \mu((V_{g})_{i})$ = V. 
Functionally, this query be written using function currying as: 
\begin{equation}\label{eqn:spt1}
    max(values_{age}(out_{knows}(has_{name=marko}(V_{g})))))
\end{equation}

\end{example}
The terms out$_{knows}$, values$_{age}$ and has$_{name}$ are the single-step Gremlin operations/traversals. 
In~\cite{rodriguezN11grptrvslpttrn}, Rodriguez presents the itemisation of such single-step traversals which can be used to represent a complex path traversal.
Thus, as described earlier, through functional composition and currying one can represent a graph traversal of random length. 
If \textit{i} be the starting vertex in G, then the traversal shown in listing~\ref{lst1} can be represented as following function: 
\begin{equation}\label{eqn:fofi}
    f(i) = max(\epsilon_{age} \circ v_{in} \circ e_{lab+}^{knows} \circ e_{out} \circ \epsilon_{name+}^{marko}) \hspace{5 pt} (i)
\end{equation}
where, \textit{f} : \textit{P(V)} $\rightarrow$ \textit{P(S)}.
A detailed illustration of the single step traversals can be referred from~\cite{rodriguezN11grptrvslpttrn}.

\subsection{Evaluation of \texttt{match()}-step in Gremlin}\label{sec:gremlin_match}
The \texttt{match()}-step\footnote{\url{http://tinkerpop.apache.org/docs/3.2.3/reference/\#match-step}} evaluates the input graph patterns over a graph $G$ in a structure preserving manner binding the variables and constants to their respective values (cf. Definition~\ref{def:match}). 
We denote the evaluation of a traversal $t$ over a graph $G$, using the notation $[[t]]_{g}$ which is borrowed from~\cite{perez2006semantics,angles2008expressive}. 
When a \texttt{match()}-step is encountered by the Gremlin machine, it treats each graph pattern as an individual path traversal. 
These graph patterns are represented using \texttt{as()}-step\footnote{Meaningful names can be used as variable names for enhancing query readability.} (step-modulators\footnote{Rodriguez et al.~\cite{rodriguezN11grptrvslpttrn} refer to step modulators as `syntactic sugar' that reduce the complexity of a step's arguments by modifying the previous step.} i.e. naming variables) such as $a$, $b$, $c$, etc.), which typically mark the start (and end) of particular graph patterns or path traversals.

However, the order of execution of each graph pattern is up to the \texttt{match()}-step implementation, where the variables and path labels are local only to the current \texttt{match()}-step.
Due to this uniqueness of the Gremlin \texttt{match()}-step it is possible to: 
\begin{enumerate}
    \item treat each graph pattern individually as a single step traversal and thus, construct composite graph patterns by joining (path-joins) each of these single step traversals; 
    \item combine multiple \texttt{match()}-steps for constructing complex navigational traversals (i.e. multi-hop queries), where each composite graph pattern (from a particular \texttt{match()}-step) can be joined using the concatenative join (ref. Section~\ref{sec:bin_ops}).
\end{enumerate}

For instance, consider the GPM gremlin query as shown in Listing~\ref{lst2}.
\begin{lstlisting}[caption= {This traversal returns the names of people who created a project named 'lop' that was also created by someone who is 30 years old.}, label={lst2}, language=Java]
g.V().match( 
        __.as('a').out('created').as('b'), 
        __.as('b').has('name', 'lop'), 
        __.as('b').in('created').as('c'),
        __.as('c').has('age', 30)).select('a','c').by('name')
\end{lstlisting}


Each of the comprising four graph patterns (traversals) of the query (listing~\ref{lst2}), can be individually represented using the curried functional notation as described in Equation~\ref{eqn:spt1}. Thus, 

\begin{equation}\label{eqn:cnot_master1}
    f(i) = (e_{lab+}^{created}) \circ e_{out} \hspace{5 pt} (i); \hspace{5pt} g(i) = (\epsilon_{name+}^{lop} \circ v_{in}) \hspace{5 pt} (i); 
\end{equation} 
\begin{equation}\label{eqn:cnot_master2}
    h(i) = (e_{lab+}^{created}) \circ e_{in} \hspace{5 pt} (i); \hspace{5pt} j(i) = (\epsilon_{age+}^{30} \circ v_{in}) \hspace{5 pt} (i)
\end{equation}

The input arguments of the \texttt{match()}-step are the set of graph patterns defined above in equations~\ref{eqn:cnot_master1},\ref{eqn:cnot_master2}, which form a composite graph pattern (the final traversal ($\Psi$)).
At run-time, when a traverser enters \texttt{match()}-step, it propagates through each of these patterns guaranteeing that, for each graph pattern, all the prefix and postfix variables (i.e. "a", "b", etc) are \textit{binded} with their labelled path values. 
It is only then allowed to exit, having satisfied this condition.
In simple words, though each of these graph patterns is evaluated individually, it is made sure that at run-time, the overall structure of the composite graph pattern is preserved by mapping the path labels to declared variables.

For instance, in the query (ref. listing~\ref{lst2}), the starting vertex of $g(i)$ labelled as "b" which is the terminal vertex of $f(i)$, similarly for $h(i)$ and $j(i)$ with vertex labelled as "c".
It is therefore necessary, to keep a track of the current location of a traverser in the graph, to preserve traversal structure. 
This is achieved in Gremlin by \texttt{match()} and \texttt{bind()} functions respectively, which we outline next.

The evaluation of an input graph pattern/traversal in Gremlin is taken care by two functions: 
\begin{enumerate}
    \item the recursively defined \texttt{match()} function- which evaluates each constituting graph pattern and keeps a track of the traversers location in the graph (i.e. path history), and, 
    \item the \texttt{bind()} function- which maps the declared variables (elements and keys) to their respective values. 
\end{enumerate}
Using equations~(\ref{eqn:cnot_master1},~\ref{eqn:cnot_master2}) (curried functional form of path traversals) in the recursive definition of \texttt{match()} by~\cite{rodriguez2015trvslmchnlang}, we have: 
\begin{equation}\label{eqn:match_master}
\text{[[}t\text{]]}_{g} = \left \{ 
\begin{tabular}{ll}
$\text{[[}bind_{b}(\texttt{f}(t_{{\Delta}_{a}(t)}\wedge\Delta_{m1}))\text{]]}_{g}$ & \hspace{5pt} : $\Delta_{a} \neq \phi = \Delta_{m1}$  \\
$\text{[[}\texttt{g}(t_{{\Delta}_{b}(t)}\wedge\Delta_{m2})\text{]]}_{g}$ & \hspace{5pt} : $\Delta_{b} \neq \phi = \Delta_{m2}$  \\
$\text{[[}bind_{c}(\texttt{h}(t_{{\Delta}_{b}(t)}\wedge\Delta_{m3}))\text{]]}_{g}$ & \hspace{5pt} : $\Delta_{b} \neq \phi = \Delta_{m3}$  \\
$\text{[[}\texttt{j}(t_{{\Delta}_{a}(t)}\wedge\Delta_{m4})\text{]]}_{g}$ & \hspace{5pt} : $\Delta_{c} \neq \phi = \Delta_{m4}$ \\
$t$ & \hspace{5pt} : otherwise,
\end{tabular} \right. 
\end{equation}
where, $t_{\Delta_{a}}$(t) is the labelled path of traverser \textit{t}. 
A path (ref. definition~\ref{def:path}) is labelled "a" via the step-modulator \texttt{.as()}, of the traverser in the current traversal ($\Psi$);
$\Delta_{m1}$, $\Delta_{m2}$, $\Delta_{m3}$ are hidden path labels which are appended to the traversers labelled path for ensuring that each pattern is executed only once; and bind$_{x}$(t) is defined as: 
\begin{equation}
bind_{x}(t) = \left \{ 
\begin{tabular}{ll}
$t_{{\Delta}_{x}}(t) = \mu(t)$ & \hspace{5pt} : $\Delta_{x}(t) = \phi$ \\
$t$ & \hspace{5pt} : $\Delta_{x}(t) = \mu'(t)$ \\
$\phi$ & \hspace{5pt} : otherwise. 
\end{tabular} \right.
\end{equation}
where $\mu'$: T $\rightarrow$ U, is a function that maps a traverser to its current location in the graph G (e.g., v $\in$ V, V $\in$ U)~\cite{rodriguez2015trvslmchnlang}. It ($\mu'$) can be perceived analogues to $\mu$ defined in definition~\ref{def:pg}, which maps elements and keys to values in G, however in later case the value is the location of a traverser in G.

\section{Mapping Gremlin GPM traversals to Graph Algebra}\label{sec:algorithm}  

\begin{figure}[t!]
\begin{center}
\includegraphics[width=\textwidth]{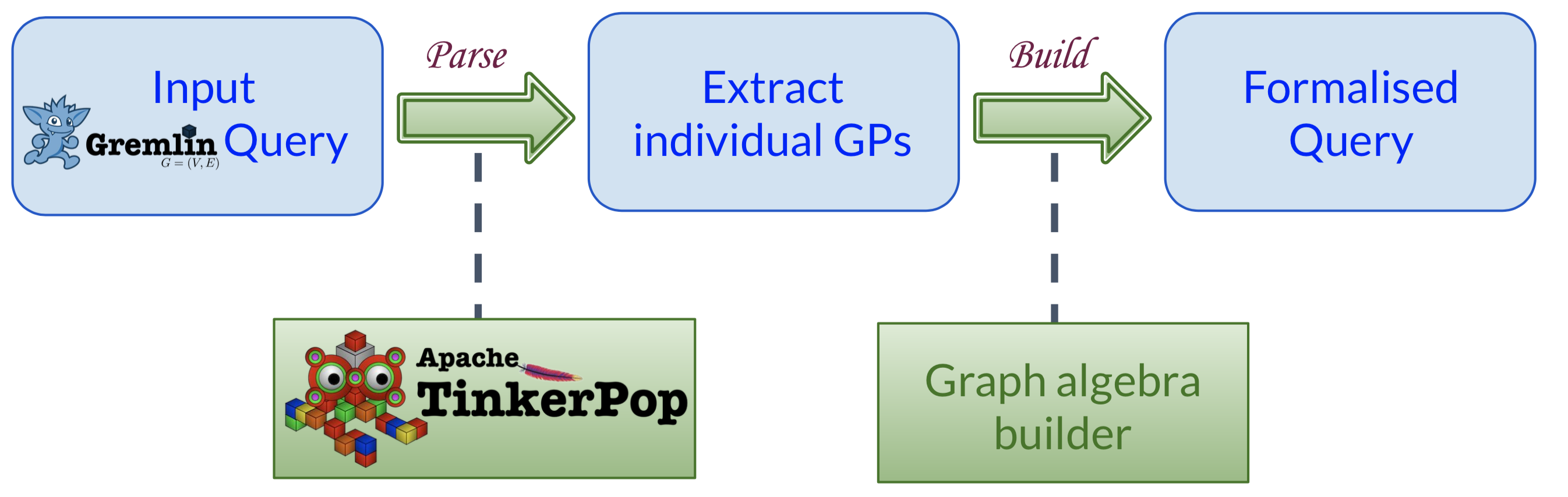}
\end{center}  
\caption{Conceptual architecture for formalizing a Gremlin traversal using graph relation algebra.} 
\label{fig:grem_formalization}
\end{figure}

In this section, we present a mapping algorithm for encoding a given Gremlin pattern-matching traversal, relational graph algebra. 
Figure~\ref{fig:grem_formalization}, describes the conceptual architecture for formalizing Gremlin traversals.
We follow a bottom-up approach in order to construct the relational graph algebra based on the traversal.

\subsection{Mapping Gremlin traversals to Graph Algebra}

\begin{enumerate}
    \item The input query is parsed and its constituent individual graph patterns are extracted from the \texttt{match()}-step and the optional \texttt{where()} step. {\color{red} \texttt{/* Parse step}}
    \item For each single graph patterns (single path traversals) in the query, we first construct the curried functional form 
    (ref. equation~\ref{eqn:spt1}).
    \item We then map the get-vertices/get-edges operator for the encountered g.V()/g.E() step (i.e. to the first graph pattern) respectively. {\color{red} \texttt{/* Steps 2-12 are the Build steps}}
    \item Append a \textit{traverse-operator} to all the respective in-coming and outgoing edge traversals for each, that appear inside the \texttt{match()}-step.
    \item Append a \textit{property-filter} operator to all the respective \texttt{has()} and \texttt{values()} steps based on the \texttt{match()}-step.
    \item Multiple \texttt{match()} steps can be connected processed using the concatenative join operator.
    \item Append a \textit{selection} operator, if the \texttt{match()} step is succeeded by a \texttt{where()} step (this is an optional in gremlin queries).
    \item Append a \textit{projection} operator, if \texttt{select()}-step is declared with a \texttt{match()} or \texttt{where()} step.
    \item Append a \textit{deduplication} operator, based on whether the \texttt{dedup()} step is declared after the \texttt{select()} step.
    \item Append a \textit{sorting} operator, if the \texttt{order()} step with an optional \texttt{by()} modifier is declared after the \texttt{select()} step.
    \item Append a \textit{grouping} operator, if the \texttt{group()} step with an optional \texttt{by()} modifier is declared after the \texttt{select()} step.
    \item Map the \textit{union} operator if the query contains a \texttt{union()}-step\footnote{It is not a common practice to use a union()-step in Gremlin GPM traversals, as multiple \texttt{match()}-steps in conjunction with \texttt{where()}-steps can be used as per required (the ad infinitium style of traversing~\cite{rodriguez2015trvslmchnlang}).}. \textit{Union} is technically a binary operator,
    however, a union of multiple patterns can be constructed using a left deep join tree representation. 
\end{enumerate}

Next we present three examples of gremlin traversals which have been formalised using the proposed graph relational algebra. 
\paragraph{Example Query 1.}
The sample query as shown in listing~\ref{lst2}, can be formalized as:  
\begin{equation}
\dagger_{name}\bigg(\Pi_{a,c}\Big(\llbracket \underbrace{\sigma_{age=30}^{c}  \downarrow_{b}^{c}{[}created{]}  \sigma_{name=lop}^{b} \uparrow_{a}^{b}{[}created{]}(V_{g})}_\text{\color{red}t}\rrbracket_{g}\Big)\bigg)
\end{equation}

\paragraph{Example Query 2.}
Consider the following gremlin traversal shown in listing~\ref{lst41} below:
\begin{lstlisting}[caption= {This traversal returns the list all the persons in the ascending order of the age.}, label={lst41}, language=Java]
g.V().match( 
    __.as('a').hasLabel('person').values('age').as('b')).select('b').order().by(asc)
\end{lstlisting}

The gremlin traversal shown above (Listing~\ref{lst41}) can be formalized as follows:  
\begin{equation}
\Re_{\Uparrow_b}\bigg(\Pi_{b}\llbracket \underbrace{\sigma_{age}^{b} \sigma_{label=person}^{a}(V_{g})}_{\text{\color{red}t}}\rrbracket_{g}\bigg)
\end{equation}

\paragraph{Example Query 3.}
Consider the following gremlin traversal shown in listing~\ref{lst51} below,
\begin{lstlisting}[caption= {This traversal returns the list of all the people who have collaboratively created a software.}, label={lst51}, language=Java]
g.V().union( 
 __.match( __.as('a').out('created').as('c')),
 __.match( __.as('b').out('created').as('c'))).select('a','c')
\end{lstlisting}

The gremlin traversal shown above (Listing~\ref{lst51}) can be formalized as follows:  
\begin{equation}
\Pi_{b,c}\bigg(\llbracket \underbrace{\uparrow_{a}^{c}{[}created{]}(V_{g})\rrbracket_{g} \uplus  \llbracket  \uparrow_{b}^{c}{[}created{]}(V_{g})}_{\text{\color{red}$t = t_{1} \uplus t_{2}$}}\rrbracket_{g}\bigg)
\end{equation}

\paragraph{Optimizations.}
The Gremlin graph traversal machine inherently offers a wide variety of machine and query level optimizations as traversal strategies including 
(i) query rewriting (decoration), 
(ii) traversal optimization, 
(iii) vendor optimization (using byte-code for gremlin language variants), 
(iv) finalization, and 
(v) verification. 
We will not go into the specific details, rather we point the interested reader to (Section 4, page 6-7 of~\cite{rodriguez2015trvslmchnlang}). \\

\paragraph{Limitations.}\label{sec:limitations}
We do not cover the complete Gremlin language, as clarified earlier, we strictly focus on formalizing the GPM (declarative) construct. 
In the declarative construct of Gremlin, in this work, we only focus on covering the core functionality of pattern-based graph traversing. 
We do not formalize mutating and specific graph-based task (such as, centrality, eccentricity, etc) traversals and data manipulation operators, e.g., \texttt{addVertex()}, \texttt{addEdge()}, \texttt{addProperty()}, \texttt{aggregate()} operators.

\section{Related Work}\label{sec:relwork}

This section summarizes work towards formalizing query algebras for different data models and corresponding query languages.

\paragraph{Property graphs.} 
The Property Graph data model is one of the popular graph data models that provides a rich set of features for the user to model domain-specific real world data. 
Various query languages have been proposed over the years for querying over PGs. 
The \textit{Cypher} query language\footnote{\url{https://neo4j.com/developer/cypher-query-language/}}, native to the Neo4J PG database, is a high-level declarative pattern matching-based graph query language. 
The developers of Neo4J \& Cypher strive at standardizing Cypher by providing open formal specification via the OpenCypher\footnote{\url{http://www.opencypher.org/}} project~\cite{szarnyasopencypher}. 
One of the limitations of Cypher is that it misses certain graph querying functionality such as the support for regular path queries and graph construction.
\textit{PGQL}~\cite{van2016pgql}, is an SQL-like syntax based graph query language for the PG data model. 
Albeit being able to overcome the limitations of Cypher, and lure the SQL community with its SQL-like syntax support, PGQL  lacks standardization and support by database technology vendors.

\paragraph{RDF:} 
The Resource Description Framework (RDF), is another graph data model, popular in the semantic web domain. 
In RDF the data (i.e., entity descriptions) are stored as triples, similar to the node-edge formalism in PGs. 
SPARQL~\cite{prud2006sparql}, the query language for RDF triple stores, is a Cypher-like declarative GPM query language for querying RDF graphs. 
SPARQL is a W3C standard and its query algebra has been formally described in works such as~\cite{perez2006semantics,angles2008expressive}. 
Moreover, multiset semantics have also been formalized by~\cite{AnglesG16Multiset,schmidt2010foundations}.

\paragraph{SQL:} 
Relational databases cater rather limited support for executing graph queries. However, certain databases such as PostgreSQL allow the execution of recursive queries. The SAP HANA Graph Scale-Out Extension prototype (GSE), using a SQL-type language proposed by~\cite{krause2016sql}, supports modelling high level graph queries.

\section{Conclusion and Outlook}\label{sec:conclusion}
In this extended paper, we presented the initial efforts on formalizing GPM traversals, which is a subset of the Gremlin traversal language and machine. 
Since, Gremlin is both a query language and a machine, it enables the graph (pattern-match) traversals to be represented in any formal query language, given that it supports function composition and nesting. 
Our current work provides a theoretical foundation to leverage this advantage of Gremlin for querying graphs on various graph engines.
Furthermore, our current work lays the foundation for supporting query interoperability by allowing translation of SPARQL queries to Gremlin GPM traversals~\cite{thakkar2018stitch,91512}, using the proposed mapping, in order to: \textit{(i)} leverage the advantage of using graph traversal for query property graphs, and \textit{(ii)} enable the use of SPARQL query language (the defacto standard of RDF stores) on top of other OLAP-based graph processors and OLTP-based graph engines~\cite{thakkar2018stitch,91512}.
This enables the semantic web community to reap best of both worlds, i.e., accessing a plethora of graph data management systems without the obligation of embracing a new graph query language. 
As the near future work, we aim to address the limitations pointed out in Section~\ref{sec:limitations}. 
\vspace{20pt}

\hspace{-15pt}\textbf{Acknowledgement}
\begin{table}[ht]
    \begin{tabular}{l p{10cm}}
     \includegraphics[scale=0.17]{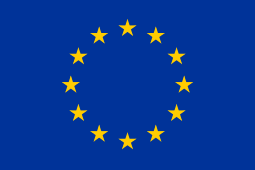} & \vspace{-25pt} This work is supported by the funding received from EU-H2020 Framework Programme under grant agreement No. 780732 (BOOST4.0)
    \end{tabular}
    \label{tab:my_label}
\end{table} 

\bibliographystyle{abbrv}
\bibliography{ref}

\end{document}